\documentclass[aps,twocolumn,groupedaddress,showpacs,floatfix]{revtex4}

\begin{document}
\bibliographystyle{apsrev}

\title{
A mean field approach for string condensed states
}

\author{Michael Levin}
\author{Xiao-Gang Wen}
\homepage{http://dao.mit.edu/~wen}
\affiliation{Department of Physics, Massachusetts Institute of Technology,
Cambridge, Massachusetts 02139}
\date{Mar. 2004}

\begin{abstract}
We describe a mean field technique for quantum string (or dimer) models. 
Unlike traditional mean field approaches, the method is general enough to 
include string condensed phases in addition to the usual symmetry 
breaking phases. Thus, it can be used to study phases and phases transitions
beyond Landau's symmetry breaking paradigm. We demonstrate the technique 
with a simple example: the spin-$1$ XXZ model on the Kagome lattice. 
The mean field calculation predicts a number of phases and phase 
transitions, including a $z=2$ deconfined quantum critical point. 
\end{abstract}

\maketitle

\section{Introduction}
Recently, several frustrated spin systems have been discovered with 
the unusual property that their collective excitations are described by 
Maxwell's equations. \cite{Wlight,MS0204,Walight,MS0312,Wen04,HFB0404}
These light-like collective modes can be traced to 
the highly entangled nature of the ground state. In these systems, the
low energy degrees of freedom are not individual spins, but rather
string-like loops of spins. The ground state is a coherent superposition 
of many such string-like configurations - a ``string condensate." It is 
this ``string condensation" in the ground state that is responsible for 
the emergent photon - just as particle condensation is responsible for the 
phonon modes in a superfluid. \cite{Walight,LWuni,LWstrnet,LWqed} 

While this qualitative picture is relatively clear, quantitative results on
string condensation and artificial light are lacking. The above models have
only been analyzed in limiting and unrealistic cases. We do not know any
realistic systems with emergent photons. The problem is that we are missing a
good mean field approach for exotic states. Current mean field theory
approaches can only be applied to symmetry breaking states with local order
parameters. They are useless for understanding string condensed states 
which are highly entangled and have nothing to do with symmetry breaking. 

In this paper, we address this problem. We describe a mean field approach 
that can be applied to both symmetry breaking \emph{and} string condensed 
states. We hope that this technique can be used to identify conditions 
under which string condensation may occur, and to help further the 
experimental search for emergent photons and new states of matter.
  
In practice, our approach can be thought of as a mean field technique for 
quantum string (or dimer) models. This technique can be used to estimate 
the phase diagram of string (or dimer) models, to find the low 
energy dynamics of the different phases, and to analyze the phase 
transitions. It can be applied to any quantum spin system with the property 
that its low energy degrees of freedom are strings or dimers. This includes 
all the frustrated spin systems cited above. 

We demonstrate the technique with a simple example: a spin-$1$ XXZ model 
on the Kagome lattice \cite{Walight}:
\begin{align}
\label{spkag}
H &= J_1\sum_{\v I} (S^z_{\v I})^2
+J_2\sum_{\<\v I\v J\>} S^z_{\v I} S^z_{\v J} \nonumber \\
&-J_{xy}\sum_{\<\v I\v J\>} (S^x_{\v I} S^x_{\v J} +S^y_{\v I} S^y_{\v
J})
\end{align}
Here $\v I$ and $\v J$ label the sites of the Kagome
lattice, and $\sum_{\<\v I\v J\>}$ sums over all nearest neighbor sites.
This model provides a good testing ground for the method since the low 
energy dynamics of $H$ is described by a string model in the regime $J_2 
\gg J_{xy} \gg |J_1 - J_2|$. 

The mean field calculation predicts a number of interesting phases 
including string condensed phases with emergent photons. The string 
condensed phases are ultimately destroyed once instanton 
fluctuations are included, but several phases and phase transitions
remain - including a deconfined quantum critical point.

\begin{figure}[t]
\centerline{
\includegraphics[width=2.5in]{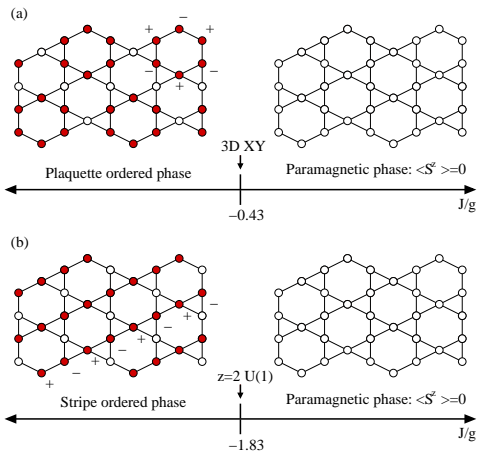}
}
\caption{
The mean field phase diagram for (a) the XXZ model (\ref{spkag}) and (b)
the XXZ model with additional next nearest neighbor interactions. The 
filled circles denote spins with $\<S^z\> \neq 
0$. The sign of $\<S^z\>$ alternates around each plaquette in the 
plaquette ordered phase and along each stripe in the stripe ordered phase. 
}
\label{mfsum}
\end{figure}

The mean field phase diagram for (\ref{spkag}) is shown in Fig. 
\ref{mfsum}a. Here, $J = \frac{9 J_{xy}^2}{J_2} + \frac{24 
J_{xy}^3}{J_2^2} + 3(J_1 - J_2)$, and $g = \frac{3 J_{xy}^3}{J_2^2}$. For 
large positive $J/g$, the system is in a paramagnetic phase with no 
broken symmetries, while for large negative $J/g$, the system is in a 
plaquette ordered phase with broken lattice and spin symmetries. The 
critical point is in the universality class of the 3D $XY$ model. This is 
similar to, but slightly different from the phase diagram obtained in 
\Ref{XM0555}. In that paper, the authors also predict a plaquette phase
at large negative $J/g$, but their candidate phase is a \emph{resonating} 
plaquette phase which has different symmetries from the \emph{frozen} 
plaquette phase shown above.

We also study the model (\ref{spkag}) with an additional second nearest 
neighbor interaction $J_{3}\sum_{\<\<\v i \v j\>\>} S^z_{\v i} S^z_{\v 
j}$, $J_3/g = 0.17$. We find a different phase diagram (Fig. 
\ref{mfsum}b). For large positive $J/g$, the system is in a paramagnetic 
phase, while for large negative $J/g$ the system is in a stripe ordered 
phase with broken rotational and spin symmetry. The mean field 
calculation predicts that the phase transition is a deconfined quantum 
critical point described by a $U(1)$ gauge theory with dynamical exponent
$z=2$. However, we cannot rule out the possibility of a first order phase 
transition - for either of the two models. 

The paper is organized as follows. In section \ref{stringpict} we 
describe the string picture for the Kagome model (\ref{spkag}). In section
\ref{mfapproach} we present the mean field approach and derive the
mean field phase diagram for (\ref{spkag}). In section \ref{lowen},
we derive the low energy dynamics in each of the phases, and in section
\ref{phasetrans} we analyze the phase transitions in the model. The
details of the mean field calculation are presented in the Appendix.

\begin{figure}[t]
\centerline{
\includegraphics[width=2.0in]{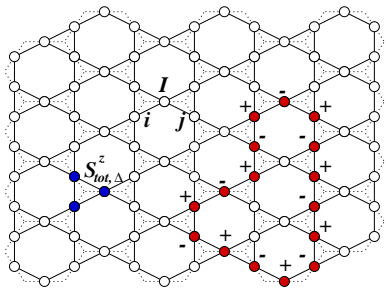}
}
\caption{
When $J_1 - J_2 = J_{xy} = 0$, the Kagome XXZ model (\ref{spkag}) 
has an extensive ground state degeneracy. All states satisfying 
$S^z_{\text{tot},\Delta} = \sum_{\v I \in \Delta} S^z_{\v I} = 0$ are 
ground states. If we view the sites of the Kagome lattice as links of a 
honeycomb lattice, then the ground states are collections of closed loops 
of alternating $S^z = \pm 1$ on the honeycomb lattice.
}
\label{Kagome2}
\end{figure}

\begin{figure}[t]
\centerline{
\includegraphics[width=2.0in]{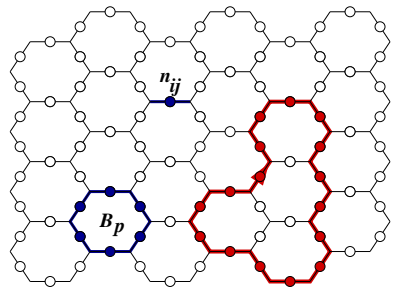}
}
\caption{
The low energy effective string Hamiltonian (\ref{Hstr}) is made up of 
two terms: an operator $n_{\v i \v j}$ which is the string occupation 
number on the link $\v i \v j$ and an operator $B_{\v p}$ which creates 
(or moves) strings along the boundary of the plaquette $\v p$.
}
\label{strmod}
\end{figure}

\section{String picture}
\label{stringpict}
\subsection{Effective string model}
We will study the XXZ model (\ref{spkag}) in the regime $J_2 \gg 
J_{xy} \gg |J_1 - J_2|$. In that case the low energy dynamics of $H$ is 
described by a string model - a close cousin of a quantum dimer model. 
\cite{Walight} To see this, note that the Hamiltonian can be rewritten as
\begin{align}
H &= \frac{J_2}{2}\sum_{\Delta} (S^z_{\text{tot},\Delta})^2 
+ (J_1 - J_2)\sum_{\v I} (S^z_{\v I})^2 \nonumber \\ 
&- J_{xy} \sum_{\<\v I \v J\>} (S^x_{\v I} S^x_{\v J} +S^y_{\v I} S^y_{\v 
J})
\end{align}
where $S^z_{\text{tot},\Delta} = \sum_{\v I \in \Delta} S^z_{\v I}$,
and $\Delta$ runs over the triangles in the Kagome lattice (see Fig. 
\ref{Kagome2}).

Suppose that $J_{xy} = J_1 -J_2 = 0$. In
that case, $H$ has an extensive ground state degeneracy: every state satisfying
$S^z_{\text{tot},\Delta}=0$ for all triangles $\Delta$, is a ground state. To 
describe these ground states, it is useful to view the sites $\v I$ of the 
Kagome lattice as the links of a honeycomb lattice - whose vertices we 
will label by $\v i$. One ground state is the state with $S^z_{\v I} = 0$ for 
all links $\v I$ of the honeycomb lattice. Another ground state can be obtained 
by alternately increasing and decreasing $S^z_{\v I}$ along a closed loop on 
the honeycomb lattice. In general all the ground states are of this form: they 
consist of collections of closed loops of alternating $S^z_{\v I} = \pm 1$ 
superimposed on a background of $S^z_{\v I} = 0$. (see Fig. 
\ref{Kagome2}). 

These states can be thought of as configurations of oriented strings on the 
honeycomb lattice. To do this precisely, we pick an $A$ and $B$ sublattice 
of the honeycomb lattice. For any link $\v I = \<\v i \v j\>$, we say 
that $\v I$ contains an oriented string pointing from $\v i \in A$ to $\v 
j \in B$ if $S^z_{\v I} = +1$, and from $\v j$ to $\v i$ if $S^z_{\v I} = -1$.
We say the link is empty if $S^z_{\v I} = 0$ (see Fig. \ref{strmod}). Then the
ground states described above are in exact correspondence with configurations
of oriented closed strings on the honeycomb lattice.

Now consider the case where $J_{xy}, J_1 - J_2$ are small but nonzero. These
terms will split the extensive degeneracy described above. The splitting can 
be described in degenerate perturbation theory by a low energy effective 
string Hamiltonian $H_{eff}$. Working to third order in $J_{xy}/J_2$ and 
assuming $J_{xy} \gg |J_1 - J_2|$ we find:
\begin{equation}
H_{eff} = \frac{J}{3}\sum_{\v i \v j} n_{\v i \v j} - 
\frac{g}{2} \sum_{\v p} (B_{\v p} + h.c) \\
\label{Hstr}
\end{equation}
where $J = \frac{9 J_{xy}^2}{J_2} + \frac{24 J_{xy}^3}{J_2^2} + 3(J_1 - 
J_2)$,
and $g = \frac{3 J_{xy}^3}{J_2^2}$. Here, the two operators, 
$n_{\v i \v j}, B_{\v p}$ are operators that act on oriented string 
states. The operator $n_{\v i \v j}$ is the string occupation number on the 
link $\v i \v j$. That is, for any string state $|X\>$,
$n_{\v i \v j} |X\> = |X\>$ if there is a string on $\<\v i \v j\>$ with
orientation $\v i \rightarrow \v j$ and 
$n_{\v i \v j} |X\> = 0$ otherwise. The operator $B_{\v p}$ 
acts on the six links along the boundary of the plaquette $\v p$. 
It's action is given by
\begin{displaymath}
B_{\v p}
\left | \bmm\includegraphics[height=0.6in]{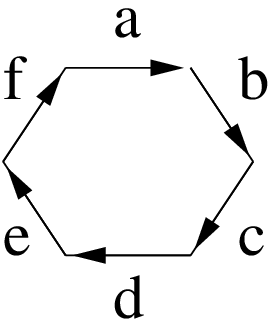}\emm \right\>
=
\left | \bmm\includegraphics[height=0.6in]{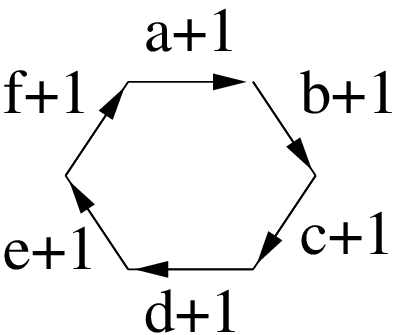}\emm \right\>
\end{displaymath}
where $a = 0$ denotes the empty link and $a = +1$ $(-1)$ denotes the link with
a string oriented clockwise (counterclockwise) with respect to $\v p$ 
(see Fig. \ref{strmod}). If any of $a,...,f = +1$, $B_{\v p}$ annihilates 
the state: $B_{\v p} |a,b,c,d,e,f\> = 0$. One can think of $B_{\v p}$ 
as being analogous to the dimer flip operators in quantum dimer models.

In the remainder of the paper, we will focus entirely on this low energy
effective string model (\ref{Hstr}). While our discussion will focus on 
strings we would like to emphasize that the results tell us about the 
physics of the simple spin model (\ref{spkag}).

\subsection{Possible phases}
The two terms in the Hamiltonian (\ref{Hstr}) have a simple physical 
interpretation in the string language. The first term, $J/3\sum_{\v i \v 
j} n_{\v i \v j}$ is a string tension which penalizes strings for being 
long. The second term, $g/2 \sum_{\v p} (B_{\v p} + h.c)$ is a string 
kinetic energy term. It makes the strings fluctuate and gives them dynamics. 

We are interested in the phase diagram of this model. We will focus on 
the case $g > 0$ for simplicity. There are three basic regimes to
consider: $J/g \gg 1$, $|J|/g \ll 1$ and $-J/g \gg 1$. 

\begin{figure}
\centerline{
\includegraphics[width=3.0in]{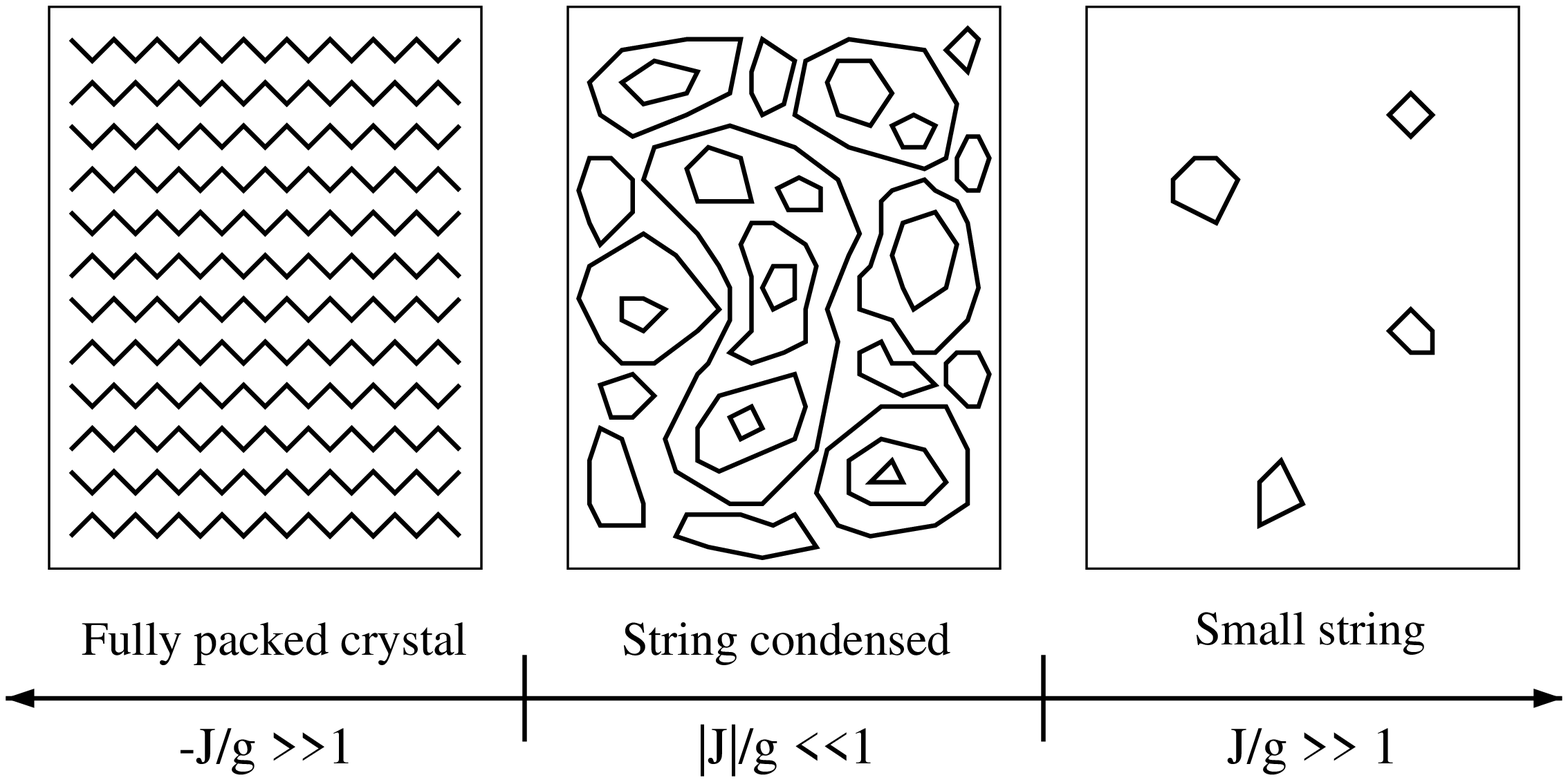}
}
\caption{
One possible phase diagram for the string model (\ref{Hstr}). The string
orientations have been omitted for clarity.
}
\label{schestrph}
\end{figure}

When $J/g \gg 1$ the string tension term dominates. In this
regime, the physics of (\ref{Hstr}) is clear. We expect the ground state
to be the vacuum state with a few small strings; the system is in
a ``small string'' phase. On the other hand, when $|J|/g \ll 1$, the string 
kinetic energy dominates. In this regime, the behavior of (\ref{Hstr}) is
less straightforward. One plausible scenario is that kinetic energy term 
favors a ground state which is a superposition of many large strings -
a ``string condensed'' state. (see Fig. \ref{schestrph}) However, it is also 
possible that the energetics favor a string crystal phase, or even a 
small string phase. In the final regime, when $-J/g \gg 1$, the 
negative string tension favors ``fully packed'' string states - states where 
every point on the honeycomb lattice is contained in a string. Again, the 
physics is not clear. It is possible that the system enters a 
fully packed string crystal phase - but it could equally well realize a 
fully packed string liquid state.

Clearly, even the qualitative phase diagram for the string model 
(\ref{Hstr}) is not obvious. There are many potentially competing phases
with very different properties. We would like to have a method for 
determining which of these phases are actually realized and for what 
parameters. It would be particularly useful to know in what region the 
string condensed phase occurs, if at all - since this phase may contain 
gapless photon-like modes. The mean field technique described below 
accomplishes this task.

\section{Mean field approach} 
\label{mfapproach}
In this section, we describe a mean field technique for quantitatively 
computing the phase diagram for the string model (\ref{Hstr}). As 
we will see later, it can also be used to derive a low energy effective 
Lagrangian for each of the phases, and to analyze the phase transitions. We 
would like to mention that similar approaches have been used to study 
lattice gauge theory. \cite{BDM8034,HW8231}

\subsection{Variational states}
Our approach is variational - we define a large class of 
variational string wave functions, and then minimize their energy 
$\<H\>$. Let us begin by describing the variational wave functions.
The wave functions $\Psi$ have a large number of variational parameters 
$\{z_{\v i \v j}\}$ indexed by the oriented links $\v i \v j$ of the 
honeycomb lattice. For each set of
$\{z_{\v i \v j}\}$, the corresponding wave function $\Psi_{\{z\}}$ is 
defined by
\begin{equation}
\Psi_{\{z\}}(X) = \prod_{\v i \v j} z_{\v i \v j}^{n_{\v i \v j}}
\label{varstate}
\end{equation}
where $n_{\v i \v j}$ is the occupation number of the oriented link 
$\v i \v j$ in the oriented string configuration $X$. We can 
see that $z_{\v i \v j}, z_{\v j \v i}$ are string fugacities on the link 
$\<\v i \v j\>$ for the two different string orientations. 

The above variational wave function (\ref{varstate}) can accommodate many 
different kinds of states - including both string crystals and string liquids.
If $z_{\v i \v j}$ is periodic, $\Psi_{\{z\}}$ is a symmetry breaking string 
crystal state; if $z_{\v i \v j}$ is constant for all $\v i \v j$, then 
$\Psi_{\{z\}}$ is a string liquid state. The variational states 
$\Psi_{\{z\}}$ can even access the two \emph{types} of 
string liquids described in the previous section - small string states and 
string condensed states - and can capture the distinction between them. 

To see this, consider a string liquid state with $z_{\v i \v j} = 
\alpha$. The properties of the state $\Psi_{\alpha}$ can be deduced from 
the properties of the classical loop gas with statistical weight
$P(X) = |\Psi_{\alpha}(X)|^2 = \alpha^{2 L(X)}$ (here $L(X)$ is the total 
length of all the loops in $X$). This loop gas is the classical $O(2)$ 
loop gas on the honeycomb lattice and has been solved exactly. It is 
known that the loop gas has two phases separated by a phase transition at 
$\alpha_c = 2^{-1/4} \approx 0.84$. \cite{DMNS8179} When $\alpha < 
\alpha_c$, the classical loop gas is in a ``small loop'' phase, where the 
typical loop size is some finite length scale $\xi$. On the other hand, when 
$\alpha > \alpha_c$, the loop gas enters a phase with large loops of 
arbitrarily large size. Intuitively, this means that the states $\Psi_{\alpha}$ 
with $\alpha < \alpha_c$ should be regarded as small string states, while the 
states with $\alpha > \alpha_c$ should be regarded as string condensed 
states. 

In the same way it is not hard to see that the wave functions 
$\Psi_{\{z\}}$ can even accommodate \emph{simultaneous} symmetry breaking order
and string condensation. Consider, for example, the case where $z_{\v i \v j}$ 
is large and nearly constant, but has a small periodic position dependence. In 
that case, $\Psi_{\{z\}}$ will exhibit both translational symmetry breaking and 
string condensation.

\subsection{Defining string condensed states}
In the preceding section, we relied on an intuitive picture of string 
condensed states. Here we make our language more precise - formally 
defining which variational states $\Psi_{\{z\}}$ we regard as string 
condensed.

To state our definition, we place the wave function $\Psi_{\{z\}}$ on a
thermodynamically large cylinder and consider the classical loop gas 
associated with $|\Psi_{\{z\}}|^2$. We define a quantity $\rho_s$ by 
\begin{equation}
\rho_{s} = \<W^2\> -\<W\>^2
\label{strconddef}
\end{equation}
where the expectation values are taken with respect to the classical 
loop gas, and $W$ is the winding number, e.g. $W$ counts the number of 
times the loops wind around the cylinder. Our definition is 
that the state $\Psi_{\{z\}}$ is string condensed if $\rho_s > 0$ and not 
string condensed if $\rho_s = 0$ in the thermodynamic limit. Thus, 
$\rho_s$ can be thought of as a (non-local) order parameter for string 
condensation. 

The motivation for this definition is twofold. First, it captures our 
intuitive picture of string condensation: states with large 
fluctuating loops (such as $\Psi_{\alpha}$, $\alpha > \alpha_c$) have 
$\rho_s > 0$ while other states (such as $\Psi_{\alpha}$, $\alpha < 
\alpha_c$) have $\rho_s = 0$. Second, it agrees with the expected low 
energy physics of string condensed states. We will see that states with 
$\rho_s > 0$ generically have a gapless linearly dispersing photon-like 
mode (neglecting instanton effects), while other states are gapped.

\begin{figure}
\centerline{
\includegraphics[width=3.0in]{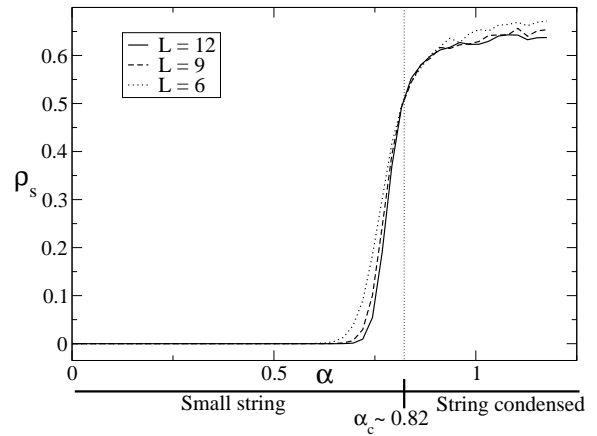}
}
\caption{
Monte Carlo results for $\rho_s$ as a function of $\alpha$, obtained 
on $6 \times 6$, $9 \times 9$ and $12 \times 12$ lattices. In the 
thermodynamic limit, $\rho_s$ vanishes for the small
string states and then jumps to a finite non-zero value for the string 
condensed states. We estimate the critical point $\alpha_c \sim 0.82$ as 
the value of $\alpha$ where the three finite size curves intersect. 
}
\label{windgraph}
\end{figure}

In addition to clarifying our discussion of string condensation,
$\rho_s$ can be used in numerical simulations to distinguish string 
condensed $\Psi_{\{z\}}$ from normal $\Psi_{\{z\}}$. As a demonstration, 
we have computed  $\rho_s$ numerically for the liquid states, $z_{\v i \v 
j} = \alpha$ and found that the transition from small string states to 
string condensed states occurs at $\alpha_c \approx 0.82$ (Fig. 
\ref{windgraph}). In this particular case this computation was 
unnecessary, since the transition point was already known analytically: 
$\alpha_c = 2^{-1/4} \approx 0.84$. However, in more complicated cases 
(such as states with non-uniform $z_{\v i \v j}$), such a numerical 
calculation would be necessary.

Finally, we would like to mention a physical interpretation of $\rho_s$
which will prove useful later. This interpretation is based on the 
duality between the classical $O(2)$ loop gas and the classical (2D) $XY$ 
model. \cite{DMNS8179} This duality maps the small loop phase of the classical 
loop gas onto the disordered phase of the $XY$ model, the large loop phase onto 
the (algebraically) ordered phase of the $XY$ model, and the transition 
at $\alpha = \alpha_c$ onto the Kosterlitz-Thouless transition. Under 
this duality, the quantity $\rho_s$ corresponds to the superfluid 
stiffness in the dual $XY$ model. This means that our previous definition 
of string condensed states can be rephrased as: $\Psi_{\{z\}}$ is string 
condensed if and only if the $XY$ model dual to the loop gas 
$|\Psi_{\{z\}}^2|$ is in the ordered phase.

\subsection{Mean field phase diagram}
The mean field phase diagram can be obtained by minimizing
the ground state energy $\<H\>$ over all states $\Psi_{\{z\}}$, and then
identifying the quantum phase associated with the minimum energy $\Psi_{\{z\}}$.
Energy expectation values can be obtained in a number of ways - we
compute them numerically using a variational Monte Carlo method. 

\begin{figure}
\centerline{
\includegraphics[width=3.0in]{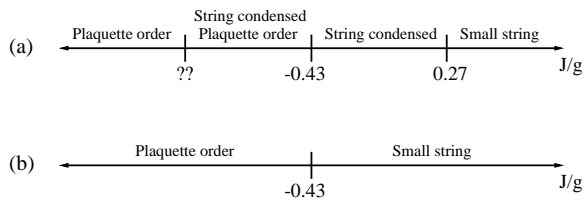}
}
\caption{
(a) The phase diagram obtained from the mean field approach.
(b) The phase diagram after instanton effects have been included. The
string condensed phases are destroyed and only two phases remain.
}
\label{hexphdiag}
\end{figure}

We have applied this technique to (\ref{Hstr}), using the energy minimization 
procedure described in Appendix \ref{minenergy}. We find the mean field phase 
diagram shown in Fig. \ref{hexphdiag}a. We find that when $J/g > 0.27$, the 
system is in a small string liquid phase. When $-0.43 < J/g < 0.27$ the system 
is in a string condensed liquid phase. When $J/g < -0.43$, the system enters a 
phase with simultaneous string condensation and plaquette order. We have not 
executed systematic numerics beyond this point but we believe that when 
$J/g$ becomes sufficiently large and negative, the string condensation
is destroyed, and the system enters a phase with plaquette order and no
string condensation. (See Appendix \ref{mfphdiag} for details on how these 
results were obtained).

\begin{figure}
\centerline{
\includegraphics[width=3.0in]{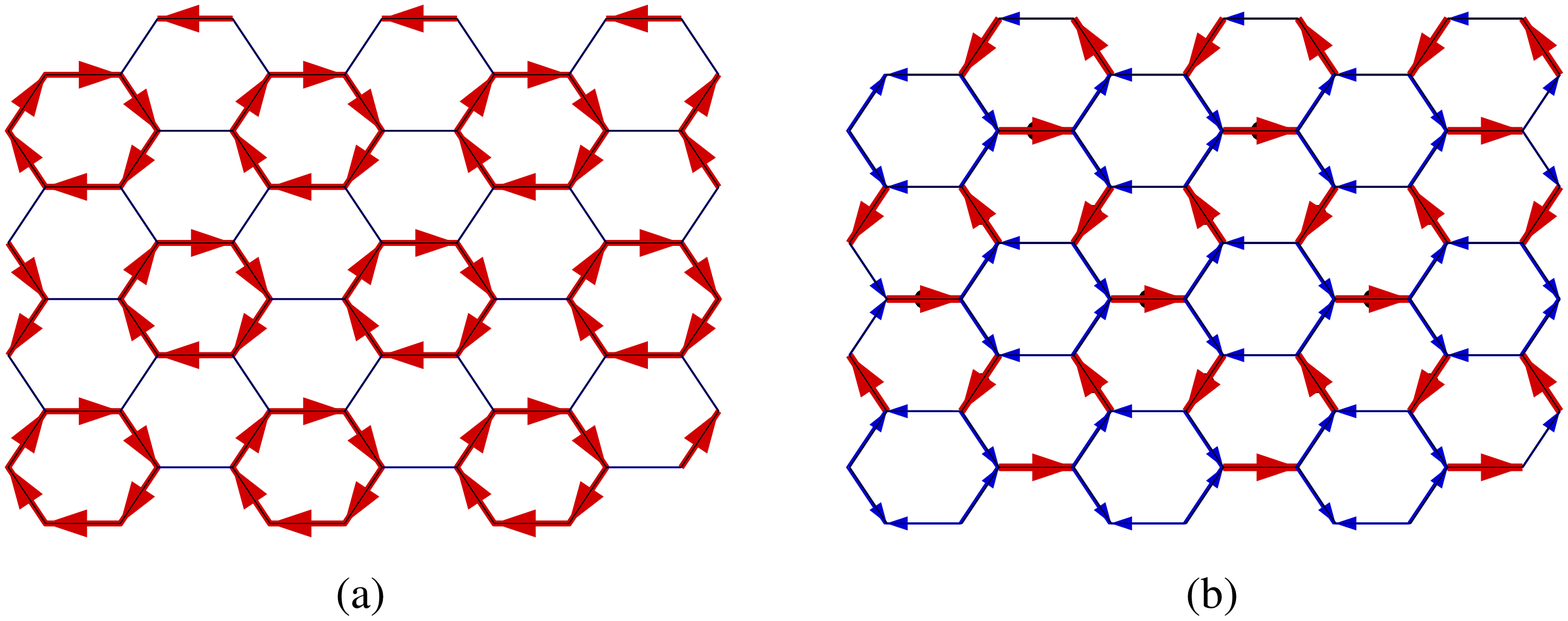}
}
\caption{
The mean field calculation predicts (a) frozen plaquette order for $J/g < 
-0.43$. Previous work \cite{XM0555} has suggested (b) resonating 
plaquette order. The two orders are shown above: the thickness
of the bonds indicates the size of
$z_{\v i \v j} \cdot z_{\v j \v i}$ while the arrows indicate the relative
size of $z_{\v i \v j}$ and $z_{\v j \v i}$. 
}
\label{frozresplaq}
\end{figure}

As we will show later, the true phase diagram is different from the
mean field phase diagram due to instanton effects. Once these are taken into
account, the two string condensed phases are destroyed, and only the
small string and plaquette ordered phases remain (Fig. \ref{hexphdiag}b).
This is similar to, but slightly different from the phase diagram obtained in 
\Ref{XM0555}. In that paper, the authors predict a \emph{resonating} 
plaquette phase (Fig. \ref{frozresplaq}b) at large negative $J/g$, while 
our mean field phase diagram predicts a \emph{frozen} plaquette phase 
(Fig. \ref{frozresplaq}a). The two phases have the same translational 
symmetry group, but different symmetries under spatial and spin rotations.

\begin{figure}
\centerline{
\includegraphics[width=3.0in]{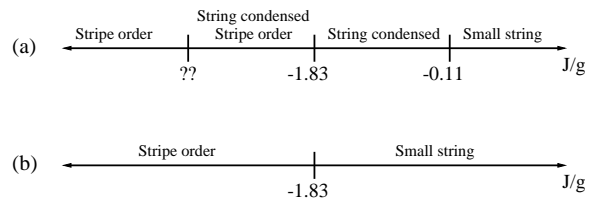}
}
\caption{
(a). The mean field phase diagram for the XXZ model with a small
second nearest neighbor interaction
$J_{3}\sum_{\<\<\v i \v j\>\>} S^z_{\v i} S^z_{\v j}$, $J_{3}/g = 0.17$.
(b). The phase diagram after instanton effects have been included.
}
\label{striphdiag}
\end{figure}

\begin{figure}
\centerline{
\includegraphics[width=1.5in]{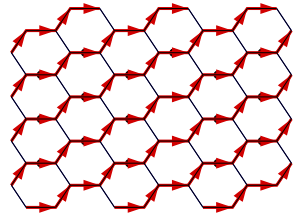}
}
\caption{
The addition of a second nearest neighbor interaction leads to a stripe 
ordered phase, depicted above. The thickness of the bonds indicates the 
size of $z_{\v i \v j} \cdot z_{\v j \v i}$ while the arrows indicate the 
relative size of $z_{\v i \v j}$ and $z_{\v j \v i}$.
}
\label{strorder}
\end{figure}

We have also computed the mean field phase diagram from the model 
(\ref{spkag}) with an additional next nearest neighbor interaction
$J_{3}\sum_{\<\<\v i \v j\>\>} S^z_{\v i} S^z_{\v j}$, $J_3/g = 0.17$. We 
find that the phase diagram shown in Fig. \ref{striphdiag}a. When 
$J/g > -0.11$, the system is in a small string liquid phase, while when 
$-1.83 < J/g < -0.11$ the system is in a string condensed liquid phase. 
When $J/g < -1.83$, the system enters a phase with simultaneous string 
condensation and stripe order (Fig. \ref{strorder}). Instanton effects 
ultimately destroy the two string condensed phases leaving only the small 
string phase and the stripe ordered phase (Fig. \ref{striphdiag}b). 
However, as we will see in section (\ref{phasetrans}), an interesting 
deconfined quantum critical point remains - the transition between the 
small string and stripe ordered phases.

\section{Low energy dynamics}
\label{lowen}
What are the low energy dynamics in each phase? From the general string 
picture, we expect (oriented) string condensed phases
to be described by compact $U(1)$ gauge theory. Therefore, we expect that 
the two string condensed phases have a gapless photon-like mode 
(neglecting instanton effects), while the other phases are gapped. But we
would like to understand this more quantitatively. We use the mean field 
method to accomplish this task. We construct a low energy effective 
Lagrangian that describes the dynamics of the string collective modes. 

The idea is based on the coherent state approach (or dynamical variational 
approach). Suppose that for some value of $J,g$, the variational ground 
state is $\Psi_{\{\bar{z}\}}$. It is useful to label this state - once 
it's been properly normalized - as $|\{\bar{z}_{\v i \v j}\}\>$. 
While $|\{\bar{z}_{\v i \v j}\}\>$ is not the exact ground state, 
our approach is based on the assumption that a good approximation to the 
ground state can be obtained by taking linear combinations of states
$|\{\bar{z}_{\v i \v j}+\delta z_{\v i \v j}\}\>$ for small 
$\delta z_{\v i \v j}$. 
We also assume that low energy excitations can be represented by such states. 
With this assumption, we can use $|\{\bar{z}_{\v i \v j}+
\delta z_{\v i \v j}\}\>$ as coherent states and use the coherent state path 
integral to calculate the low energy dynamics. In general, the Lagrangian for 
the coherent state path integral is given by
\begin{equation}
L(\delta z,\delta \dot{z}) = 
\<\{\bar{z}+\delta z(t)\}| (i \frac{d}{dt} - H) |\{\bar{z}+\delta z(t)\}\>
\label{cohstlag}
\end{equation}
where the first piece $\<\bar{z} + \delta z | i \frac{d}{dt} | \bar{z} + 
\delta z \>$ is the Berry phase term, and the second piece 
$\<\bar{z} + \delta z | H | \bar{z} +  \delta z \>$ is the usual 
Hamiltonian evolution term. 

We will focus on the simplest case, where the ground state is a string
liquid: $\bar{z}_{\v i \v j} = \alpha$ for all $\v i \v j$. It is 
convenient to parameterize the fluctuations by 
$\bar{z}_{\v i \v j}+\delta z_{\v i \v j} = \alpha e^{E_{\v i\v j} + i A_{\v 
i \v j}}$ where $E_{\v i \v j}, A_{\v i \v j}$ are arbitrary real 
numbers. Substituting this into (\ref{cohstlag}) and expanding to quadratic 
order in $E,A$ gives a Lagrangian
\begin{equation}
L = \sum_{\v i \v j, \v k \v l} (b_{\v i \v j, \v k \v l} E_{\v i \v j} 
\dot{A}_{\v k \v l} - c_{\v i \v j, \v k \v l} E_{\v i \v j} E_{\v k \v l} - 
d_{\v i \v j, \v k \v l} A_{\v i \v j} A_{\v k \v l})
\label{lattlag}
\end{equation}
Here, the constants $b_{\v i \v j, \v k \v l},c_{\v i \v j, \v k \v l}, d_{\v i 
\v j, \v k \v l}$ are given by the correlation functions
\begin{eqnarray}
\label{coeff}
b_{\v i \v j, \v k \v l} &=& -2(\<n_{\v i \v j}n_{\v k \v l}\> - 
\<n_{\v i \v j}\>\<n_{\v k \v l}\>) \\
c_{\v i \v j, \v k \v l} &=& 2(\<n_{\v i \v j} H n_{\v k \v l}\>
+\<H n_{\v i \v j} n_{\v k \v l}\>-
2\<H\>\<n_{\v i \v j}n_{\v k \v l}\>) \nonumber \\
d_{\v i \v j, \v k \v l} &=& 2(\<n_{\v i \v j} H n_{\v k \v l}\>
-\<H n_{\v i \v j} n_{\v k \v l}\>) \nonumber
\end{eqnarray}
evaluated for the string liquid state $\Psi_\alpha$.


\subsection{Gauge structure}
One might expect that by analogy with the harmonic oscillator action,
one could quantize (\ref{lattlag}) using states of the form $|\{A_{\v 
i \v j}\}\>$ as a complete orthonormal basis. However, this is not quite 
correct. The problem is with our starting point: the
coherent states $|\{z_{\v i \v j}\}\>$ are not all distinct. The 
loop structure of the string states means that
$|\{z_{\v i \v j}\}\> = |\{\tilde{z}_{\v i \v j}\}\>$ if
$\{z\}$ and $\{\tilde{z}\}$ differ by a transformation of the form
$\tilde{z}_{\v i \v j} = e^{i(\lambda_i - \lambda_j)}z_{\v i \v j}$. 

In terms of our parameterization $z_{\v i \v j} = \alpha e^{E_{\v i \v 
j}+i A_{\v i \v j}}$, this means that $|\{A_{\v i \v j}\}\> 
=|\{\tilde{A}_{\v i \v j}\}\>$ if $\tilde{A}_{\v i \v j} = A_{\v i \v j} + 
\lambda_{\v i} - \lambda_{\v j}$. Thus, the $|\{A_{\v i \v j}\}\>$ are a
many-to-one labeling of states. To obtain a true orthonormal basis,
we need to treat each equivalence class of $\{A_{\v i \v j}\}$ as a
single state.

This many-to-one labeling is identical to the $U(1)$ gauge 
structure in $U(1)$ lattice gauge theory. Thus the above action 
(\ref{lattlag}) should be regarded as a $U(1)$ gauge theory. 

\subsection{Band structure}
To solve (\ref{lattlag}), we need to find the normal modes. The
unit cell contains three links with two possible orientations, so there are
six bands. Three of these bands correspond to symmetric configurations, 
$E_{\v i \v j} = E_{\v i \v j}$, $A_{\v i \v j} = A_{\v i \v j}$ and three 
correspond to antisymmetric configurations $E_{\v i \v j} = -E_{\v i \v j}$, 
$A_{\v i \v j} = -A_{\v i \v j}$. The symmetric bands are generically gapped 
so we will focus primarily on the antisymmetric bands.

Two of the antisymmetric bands are exactly flat with $\omega_{\v k} = 0$ 
for all $\v k$. These bands correspond to pure gauge fluctuations - modes of 
the form $A_{\v i \v j} = \lambda_{\v i} - \lambda_{\v j}$. They are 
exactly flat because pure gauge fluctuations don't change the physical 
state at all. These modes don't correspond to physical degrees of 
freedom.

The only \emph{physical} low energy degree of freedom is given by the
third antisymmetric band. This band corresponds to transverse modes of $A,E$.
In the limit $\v k \rightarrow 0$, it is given by
\begin{eqnarray}
A_{\v k,\v r \v s} &=& \frac{1}{\sqrt{3Na^2}}(\v n \cdot (\v s - \v r)) 
A_{\v k} e^{i \v k \cdot \v r} 
\label {Ars} \\
E_{\v k,\v r \v s} &=& \frac{1}{\sqrt{3Na^2}}(\v n \cdot (\v s - \v r)) 
E_{\v k} e^{i \v k 
\cdot \v r} \nonumber
\end{eqnarray}
Here $\v n \perp \v k$ is a unit vector, $a$ is the lattice spacing and 
$N$ is the number of plaquettes in the lattice.
 
Substituting these expressions into the Lagrangian (\ref{lattlag}), we find
\begin{equation}
\label{lattlagfourier}
L = \sum_{\v k} (b_{\v k} E^*_{\v k} \dot{A}_{\v k} - c_{\v k} |E_{\v k}|^2
- d_{\v k} |A_{\v k}|^2)
\end{equation}
where
\begin{equation*}
b_{\v k} = \frac{1}{3Na^2}\sum_{\v r \v s, \v r' \v s'} 
b_{\v r \v s, \v r' \v s'} (\v n \cdot(\v s- \v r)) 
(\v n \cdot (\v s' - \v r')) e^{i \v k \cdot (\v r' - \v r)}
\end{equation*}
and
$c_{\v k}$, $d_{\v k}$ are defined similarly. The dispersion can now be
easily obtained: $\omega_{\v k}^2 = \frac{4 c_{\v k } d_{\v k}}{b_{\v k}^2}$.

\subsection{Gapless photon mode}
The above Lagrangian (\ref{lattlagfourier}) is a $U(1)$ lattice
gauge theory. Therefore, one might expect that the system always contains
a gapless photon mode.

This is not the case. We will now show that the photon mode is gapped in 
certain phases. Specifically, we will show that the mode is gapped in the 
small string phase and gapless in the string condensed phase. This is exactly 
what we expect physically based on the general string condensation picture. 
What is perhaps surprising is that this confinement/deconfinement physics 
can be captured without including the compactness of the $U(1)$ gauge 
field.  

To analyze the low energy excitations, we need to consider the limit
$\v k \rightarrow 0$. In that limit, $d_{\v k} \sim d k^2 a^2$ for some 
constant $d$. One way to see this is to use the original definition of the 
Lagrangian $L$. According to that definition, $d_k$ is proportional to
\begin{displaymath}
d_k \propto \<\alpha e^{i A_{\v k,\v r \v s}}|H|\alpha e^{i A_{\v k,\v r 
\v s}}\> - \<\alpha| H | \alpha \>
\end{displaymath}
But it's not hard to see that 
\begin{displaymath}
\<\alpha e^{i A_{\v k,\v r \v s}}|H|\alpha e^{i A_{\v k,\v r \v s}}\> -
\<\alpha| H | \alpha \> \propto \sum_{\v p} (\cos(F_{\v p}) - 1)
\end{displaymath}
where $F_{\v p} = A_{12} + A_{23} + A_{34} + A_{45} + A_{56} + A_{61}$ is the 
flux through the plaquette $\v p$. Since for small $\v k$, $\cos(F_{\v 
p})-1 \sim k^2 a^2$ we conclude that $d_{\v k} \sim d k^2 a^2$ for some 
constant $d$. 

Because $d_{\v k} \rightarrow 0$ as $\v k \rightarrow 0$, there are potentially
gapless excitations at $\v k = 0$. The presence or absence of a gap depends on 
the behavior of $b_{\v k}$ and $c_{\v k}$ at small $\v k$. To understand 
this behavior, we make use of the duality between the classical 
loop gas and the $XY$ model. \cite{DMNS8179} Under this duality the quantity 
$n_{\v r \v s} - n_{\v s \v r}$ corresponds to the boson current 
$(\v s - \v r) \cdot \v j(\v r)$ ($j(\v r) = \nabla \theta(\v r)$). This 
means that $b_{\v k}$ can be identified 
with the current-current correlator, 
$b_{\v k} \sim \<(\v n \cdot \v j_{\v k})(\v n \cdot \v j_{-\v k})\>$.

There are two regimes to consider. In the string condensed phase, 
$\alpha > \alpha_c$, the $XY$ model is in the ordered phase. In this case, 
the current-current correlator is of the form
\begin{equation}
\<j^{\mu}_{\v k} j^{\nu}_{-\v k}\> \sim \rho_s \frac{k^{\mu} k^{\nu} - k^2
\delta^{\mu\nu}}{k^2}
\label{curcorlarge}
\end{equation}
This means that $b_{\v k}$ approaches a nonzero constant value $b \sim \rho_s$ 
as $\v k \rightarrow 0$. 

On the other hand, in the small string phase, 
$\alpha < \alpha_c$, the $XY$ model is in the disordered phase. In this case,
the current-current correlator is of the form
\begin{equation}
\<j^{\mu}_{\v k} j^{\nu}_{-\v k}\> \sim \frac{k^{\mu} k^{\nu} - k^2 
\delta^{\mu\nu}}{k^2+1/\xi^2} 
\label{curcorsmall}
\end{equation}
where $\xi$ is the correlation length. This means that 
$b_{\v k} \sim k^2 \xi^2$ as $\v k \rightarrow 0$. In a similar way, we 
can derive the behavior of $c_{\v k}$ as $\v k \rightarrow 0$. Note that 
\begin{equation}
c_{\v r \v s, \v t \v u}-d_{\v r \v s, \v t \v u} = 4(\<H \delta_{\v r \v s} 
\delta_{\v t \v u}\> - \<H\>\<\delta_{\v r \v s}\delta_{\v t \v u}\>) 
\end{equation}
The Fourier transform of the left hand side is $c_{\v k} - d_{\v k}$. The
Fourier transform of the right hand side can be expanded in terms of current
operators as
\begin{equation}
\<\sum_{\v q} j^{\lambda}_{- \v q} j^{\lambda}_{\v q} j^{\mu}_{-\v k} 
j^{\nu}_{ \v k}\> -
 \<\sum_{\v q} j^{\lambda}_{- \v q} j^{\lambda}_{\v q}\>\< j^{\mu}_{-\v k} 
j^{\nu}_{ \v k}\> + ...
\end{equation}
By Wick's theorem, this expression is of the same order as $b_{\v k}^2$.  
It follows that as $\v k \rightarrow 0$, $c_{\v k} \sim d_{\v k} \sim dk^2 a^2$ 
in the small string phase, and $c_{\v k}$ approaches a nonzero constant value 
$c\sim \rho_s^2$ in the string condensed phase. 

Putting our expressions for $b_{\v k}, c_{\v k}, d_{\v k}$ together, we 
can compute the dispersion relation $\omega_{\v k}$ as $\v k \rightarrow 0$. 
We find that in the small string phase, the $\v k \rightarrow 0$ excitations 
are gapped: $\omega_{\v k} \sim \frac{\sqrt{c_{\v k} d_{\v k}}}{b_{\v 
k}} \sim \frac{d a^2}{\xi^2}$. On the other hand, in the string condensed 
phase,the excitations are gapless: $\omega_{\v k} \sim \frac{a\sqrt{c d}}{b} |k|$. 
This gapless mode is the artificial photon that we expected in the string
condensed state. The photon propagates with a ``speed of light" 
$v \sim \frac{a\sqrt{c d}}{b}$. 

We can apply the same arguments to the other phases in the mean field phase
diagram. A similar calculation predicts gapped excitations in the plaquette 
ordered phase and a gapless photon mode in the phase with simultaneous 
plaquette order and string condensation.

Clearly, the presence or absence of a gap is directly related to the
behavior of the string-string correlations at small momenta or large
distances. Only in the string condensed phase, where the string-string 
correlations decay algebraically, is a gapless photon mode present.
Indeed, this connection can be made rigorous using a single mode 
approximation argument similar to that of \Ref{RK8876}. One can prove a 
``Goldstone theorem" for (oriented) string condensation which asserts 
that any string state with algebraic string-string correlations 
$\<\delta_{\v r \v s} \delta_{\v t \v u}\> \sim 1/|\v r - \v t|^{\nu}$ 
with $\nu < 2+d$ (where $d$ is the spatial dimension) has a gapless 
photon mode.

\subsection{Instanton effect}
To fully understand the low energy physics of the string condensed phase,
we need to go back to a real space description. We restrict our
attention to low energy, long wavelength fluctuations of the form 
(\ref{Ars}). For slowly varying modes like these, the Lagrangian 
(\ref{lattlagfourier}) can be written in real space as
\begin{equation}
L = b\sum_{\v i \v j} E_{\v i \v j} \dot{A}_{\v i \v j}
- c\sum_{\v i \v j} E_{\v i \v j}^2 - d\sum_{\v p} F_{\v p}^2
\label{Lapprox}
\end{equation}
where $F_{\v p} = A_{12} + A_{23} + A_{34} + A_{45} + A_{56} + A_{61}$ is the
flux though the plaquette $\v p$. Notice this is precisely the Lagrangian 
for $U(1)$ lattice gauge theory. It gives rise to light-like collective 
excitations with a speed of light $v \sim \frac{a\sqrt{cd}}{b}$.

This is exactly what we claimed earlier. However, in the preceding 
discussion we neglected an important effect. The above lattice gauge 
theory is actually a \emph{compact} $U(1)$ gauge theory: $A_{\v i \v j}$ and 
$A_{\v i \v j} + 2\pi$ represent the same state. Therefore, the magnetic 
energy term should be changed from $F_{\v p}^2$ to $-\cos(F_{\v p})$. 
This has dramatic consequences for the low energy physics. Due to the 
non-perturbative instanton effect, compact $U(1)$ gauge theory is always 
confining in $(2+1)$ dimensions. The photon mode obtains a finite gap of 
order $\Delta \sim \sqrt{\frac{v^2}{a^2 e_0^2}} e^{-K/e^2_0}$ where $e^2_0 
\sim \sqrt{c/db^2}$ is the dimensionless gauge coupling and 
$K$ is a dimensionless constant of order $1$. \cite{P7582}

Physically, this means that fluctuations, in particular instanton 
fluctuations, modify the mean field phase diagram derived 
earlier. The instanton fluctuations prevent the strings from obtaining an
infinite correlation length, and thus destabilize the string condensed
phase. By similar reasoning, we expect the fluctuations to
destroy the phase with simultaneous string condensation and resonating
plaquette order. Thus, once we take instanton fluctuations into account, all
that remains are two phases: the small string phase and the 
plaquette phase (Fig. \ref{hexphdiag}(b)).

While the instanton fluctuations dramatically alter our phase diagram 
in $(2+1)$ dimensions, we expect that a similar mean field analysis in $(3+1)$ 
dimensions to be much more stable. In $(3+1)$ dimensions, the only effect of
quantum fluctuations (such as monopole fluctuations) is to modify the 
position of the string condensation phase transition. Therefore, our mean 
field analysis may be most useful in this context. The example 
described in this paper may be viewed as a warm up for such a three 
dimensional calculation.

\section{Nature of the phase transitions}
\label{phasetrans}
It is natural to wonder what the mean field approach can tell us about 
the phase transitions. In this section, we discuss this issue, focusing 
on phase transitions between the string condensed phase and adjoining 
phases.

\begin{figure}
\centerline{
\includegraphics[width=3.0in]{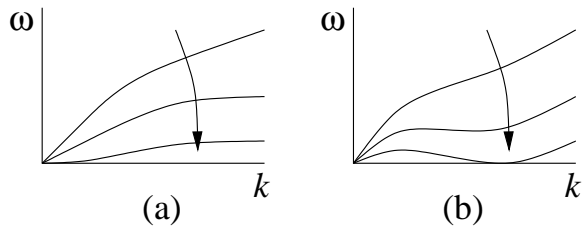}
}
\caption{
Three types of phase transitions between string
condensed phases and other phases are allowed within mean field theory.
Two are shown above. The first (a) occurs when the photon mode at $\v k = 
0$ becomes unstable. The second (b) occurs when a photon mode at some $\v 
k \neq 0$ becomes unstable.
}
\label{disp}
\end{figure}

Within mean field theory, there are three basic types of transitions that
can occur between a string condensed phase and a neighboring phase:
(1) confinement/deconfinement transitions, (2) transitions resulting from 
an instability of a $\v k \neq 0$ photon mode (Fig. \ref{disp}b), and (3)
transitions resulting from an instability at $\v k = 0$ 
(Fig. \ref{disp}a). Examples of all three of these transitions can be found in 
the simple spin-$1$ XXZ model. The third type of transition is 
particularly interesting since it is described by a $z=2$ 
deconfined quantum critical point.

\subsection{Confinement/deconfinement transition}
An example of a confinement/deconfinement transition is given by the 
critical point separating the small string and string condensed phases. 
According to mean field theory, this transition is described by a 
singularity in the variational ground state $\Psi_{\alpha}$. When 
$J/g$ is large and positive, the energetics favor a small 
string state with $\alpha < \alpha_c$. On the other hand, when $J/g$ is 
smaller, the energetics favor a string condensed state with $\alpha > 
\alpha_c$. The critical point occurs when the energetically favored 
$\alpha$ tunes through the critical value $\alpha_c$.

The behavior of the low energy excitations near the critical point can
be derived from the effective Lagrangian (\ref{lattlagfourier}). 
First, consider approaching the critical point from the small string side. 
On the small string side, the dispersion is linear in $k$ for $k \gg a/\xi^2$, 
and levels off to a gap proportional to $a^2/\xi^2$ when $k \ll 
a/\xi^2$. As we approach the critical point, the linear dispersion extends to 
longer and longer length scales and the gap goes to zero - resulting in a 
gapless photon mode.

The behavior on the string condensed side of the critical point is even 
simpler. In that case, there is no length scale other than the lattice 
spacing $a$. For all $k \ll 1/a$ the dispersion is linear in $k$. As we approach
the transition, both $b$ ($\sim \rho_s$) and $c$ ($\sim \rho_s^2$) approach 
finite non-zero values. Thus, the transition occurs at a finite 
photon velocity $v$ and a gauge coupling $e_{0}^2 \sim 1$. 

This mean field picture may capture some of the qualitative features
of the small string/string condensed phase transitions, but it is 
almost certainly incorrect when it comes to a quantitative description. 
A major source of suspicion is that the mean field phase 
transition originates from a singularity in the variational wave 
function $\Psi_{\alpha}$ itself, not from the 
fluctuations about this state. As a result the mean field exponents for 
the $(2+1)$ dimensional system come from a critical points in 
$(2+0)$ dimensions. It seems unlikely that this is the correct 
quantitative description of the critical point. 

This issue is not specific to two dimensions. If the mean field technique 
is applied to a small string/string condensed phase transition in $(3+1)$
one finds that the mean field exponents come from a critical point
in $(3+0)$ dimensions (the 3D $XY$ model). Again, this seems incorrect. 

The problem is that when we derived the Lagrangian (\ref{Lapprox}) we 
neglected higher order fluctuations, in particular the
instanton (or monopole) fluctuations of the $U(1)$ gauge field. These 
fluctuations can completely change the mean field phase transition. To 
include them we have to treat the $U(1)$ gauge field
as a compact gauge field, replacing the $F_{\v p}^2$ term in 
(\ref{Lapprox}) by $-\cos(F_{\v p})$. 

Taking these fluctuations into account gives a more accurate picture of 
the small string/string condensed critical point. In 
$(2+1)$ dimensions, instanton fluctuations destroy the string condensed 
phase altogether, and there is no phase transition at all. On the other 
hand, in $(3+1)$ dimensions, monopole fluctuations give a completely 
different mechanism for a small string/string condensed phase transition -
namely monopole condensation. This monopole mediated transition 
(which is known to be weakly first order \cite{CW7388,HLM7492}) will 
likely preempt the (unphysical) mean field critical point.

\subsection{Transition via instability at $\v k \neq 0$}
\begin{figure}
\centerline{
\includegraphics[width=1.0in]{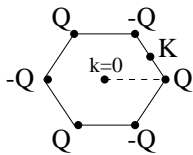}
}
\caption{
The Brillouin zone for the honeycomb lattice. The instability to plaquette
order occurs at modes (\ref{unstabmode}) with wave vectors
$\v k = \pm \v Q$.
}
\label{brillouin}
\end{figure}

The $S=1$ XXZ model also contains an example of a transition resulting
from an instability of a $\v k \neq 0$ photon mode (Fig. \ref{disp}b). This 
example occurs at the critical point separating the string condensed state 
from the state with simultaneous plaquette order and string condensation.
In the mean field picture, this transition occurs when the mode at
$\v k = \pm Q$ becomes unstable (Fig. \ref{disp12}, \ref{brillouin}). The 
critical point is described by tuning one of the coefficients $c_{\v 
Q}$ in (\ref{lattlagfourier}) through $0$ (Fig. \ref{stab123}). When $c_{\v 
Q} > 0$ the ground state is a string condensed liquid state. When $c_{\v 
Q} < 0$, nonzero $E_{\pm\v Q}$ is energetically favorable, and the ground
state acquires plaquette order (in addition to the string condensation).

\begin{figure}
\centerline{
\includegraphics[width=3.0in]{disp12.eps}
}
\caption{
The dispersion $\omega_{\v k}$ of the photon-like mode for
different values of $J/g$. As $J/g$ approaches the critical point $(J/g)_c
= -0.43$, the energy of the $\v k = \v Q$ photon goes to $0$. Thus, this
critical point is an example of the class of phase transitions depicted in
Fig. \ref{disp}b.
}
\label{disp12}
\end{figure}

Again, we need to include fluctuations to obtain the correct physics.
Instanton fluctuations gap the photon modes, destroying the string 
condensation on both sides of the transition. The result is that the 
transition is actually a simple symmetry breaking transition -between a 
small string state and a plaquette crystal state (Fig. \ref{hexphdiag}b).

To obtain the critical theory, note that due to the instanton 
fluctuations, the only low lying modes are those with $\v k \approx \pm \v 
Q$. These modes can be parameterized as
\begin{eqnarray}
E_{\v r \v s} &=& \zeta(\v r) E_{+\v Q,\v r \v s} + \zeta^*(\v r) 
E_{-\v Q,\v r \v s} \\ 
A_{\v r \v s} &=& \theta(\v r) A_{+\v Q,\v r \v s} + \theta^*(\v r) 
A_{-\v Q,\v r \v s}
\end{eqnarray}
where $\zeta(\v r),\theta(\v r)$ vary slowly on the scale of the lattice 
spacing. Substituting these expressions into the Lagrangian 
(\ref{lattlag}), and integrating out the $\theta$ field gives an action of
the form
\begin{equation}
 \cL=\frac{\rho}{2} \left(|\prt_t \zeta|^2
-v^2|\prt_x \zeta|^2 \right)
-A|\zeta|^2
\end{equation}
where $A \propto c_{\v Q}$. 

This is still not quite the correct critical theory. To get the full 
theory, we need to take fluctuations in $\zeta$ into account by including 
higher order terms. Following the analysis preceding \Eq{genzeta}, we 
include the most general terms consistent with the lattice symmetries:
\begin{equation}
 \cL=\frac{\rho}{2} \left(|\prt_t \zeta|^2
-v^2|\prt_x \zeta|^2 \right)
-A|\zeta|^2 - B|\zeta|^4 - C(\zeta^6 + \text{h.c.})
\end{equation}
The result is the critical theory for a $Z_6$ symmetry breaking 
transition - not surprising, given the symmetries of the two phases.
At the transition, the $\zeta^6$ term is irrelevant and the 
critical point is in the universality class of the 3D $XY$ model. 

We would like to mention that there is another possibility that we cannot 
rule out: the phase transition could be \emph{first order}. Because we
implemented the restricted minimization procedure described in Appendix
\ref{minenergy} we cannot resolve this question. However, in principle
the mean field technique can address this issue. One simply needs to 
use a more general minimization procedure.

\subsection{Transition via instability at $\v k = 0$}
The third type of transition - where a photon mode at $\v k = 0$ becomes
unstable 
(Fig. \ref{disp}a) - does not occur in the nearest neighbor XXZ spin model 
(\ref{spkag}). However, with an additional second nearest neighbor 
interaction 
$J_{3}\sum_{\<\<\v i \v j\>\>} S^z_{\v i} S^z_{\v j}$, such a transition 
does occur.

\begin{figure}
\centerline{
\includegraphics[width=3.0in]{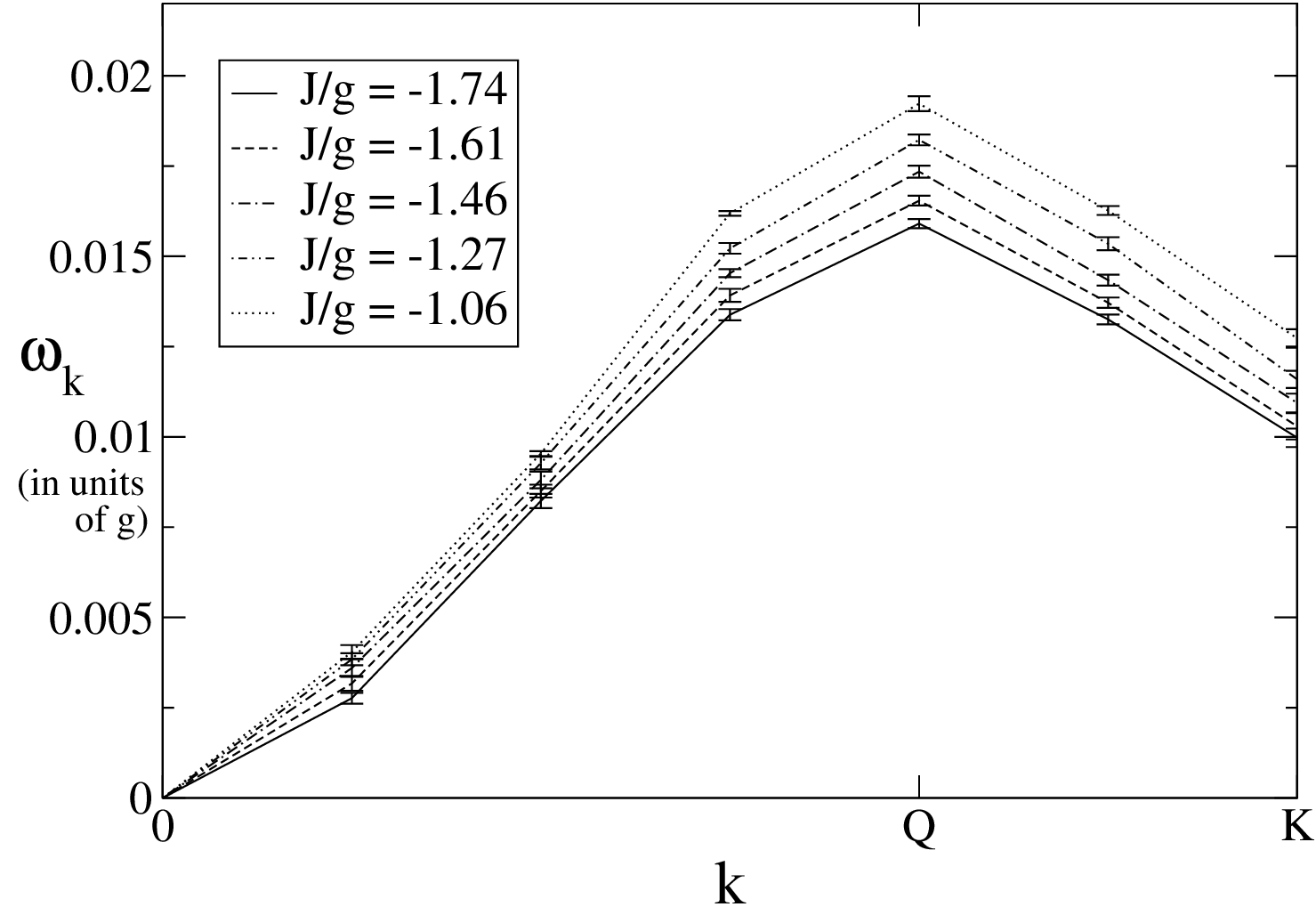}
}
\caption{
The dispersion $\omega_{\v k}$ of the photon-like mode for different values
of $J/g$ and $J_3/g = 0.17$. As $J/g$ approaches the critical point 
$(J/g)_c = -1.83$, the velocity of the photon at $\v k = 0$ goes to $0$. 
Thus, this critical point is an example of the class of phase transitions 
depicted in Fig. \ref{disp}a.
}
\label{dispstr12}
\end{figure}

With the addition of the second nearest neighbor term, the mean field phase 
diagram changes to the one shown in 
Fig. \ref{striphdiag}(a). The phase adjacent to the string condensed
phase is not plaquette ordered, but instead is stripe ordered 
(Fig. \ref{strorder}) with simultaneous string condensation. 
The transition occurs when the $\v k = 0$ mode
\begin{eqnarray}
E_{\v r \v s} &=& \v n \cdot (\v s - \v r) \\
A_{\v r \v s} &=& 0 \nonumber
\end{eqnarray}
becomes unstable (Fig. \ref{dispstr12}). The mean field critical theory is 
described by tuning the coefficient $c = \lim_{k \rightarrow 0} c_{\v k}$ 
through $0$ in the effective Lagrangian (\ref{lattlagfourier}).

It is useful to rewrite this theory in real
space. Since we are only interested in the low lying 
modes with $\v k$ small, we can make the approximation $b_{\v k} \approx b$,
$c_{\v k} \approx c + e k^2$, $d_{\v k} \approx d k^2$. Going to a real 
space, continuum description gives a Lagrangian of the form
\begin{equation}
\label{z2U1}
\cL= b' \v{E} \dot{\v{A}} - c' \v{E}^2 - d' (\nabla \times \v{A})^2 - 
e'(\nabla \times \v{E})^2
\end{equation}
where $b',c',d',e'$ are the renormalized continuum coefficients 
corresponding to $b,c,d,e$. The transition is described 
by tuning $c'$ through $0$. The critical theory is thus a 
Rokhsar-Kivelson type theory described by a $U(1)$ gauge theory with
dynamical exponent $z=2$.

The above is the mean field critical theory. In principle, we need to 
consider fluctuations to get the full critical theory. In particular, we 
need to include instanton fluctuations, replacing the $(\nabla \times 
\v{A})^2$ term with $\cos(\nabla \times \v{A})$. We also need to include 
higher order terms in $\v E, \v A$ (which are particularly important in 
the stripe ($c' < 0$) phase). The most general terms consistent with the 
lattice symmetry are given by
\begin{equation}
\cL' = -f' \v{E}^4 - g'((\v E \cdot \v n_1) (\v E\cdot \v n_2)(\v E\cdot 
\v n_3))^2 + ...
\end{equation} 
where $\v n_1, \v n_2, \v n_3$ are unit vectors along the three lattice
directions.

These fluctuations have an important effect on both phases. The instantons 
destroy the string condensation on both sides of transition. The result is
that the transition is actually a simple $Z_6$ symmetry breaking 
transition between a small string phase and a stripe phase (Fig. 
\ref{striphdiag}(b)). The higher order terms $\cL'$ are also important - 
determining the ultimate form of the ordering in the stripe phase.

However, at the critical point, both the instantons and the higher order 
terms $\cL'$ are \emph{irrelevant} for $e'$ sufficiently small. The 
critical theory is therefore described by the simple mean field Lagrangian 
(\ref{z2U1}). This result follows from the analysis in \Ref{FHM0353}. In 
that paper, the authors analyzed the $z=2$ critical point (\ref{z2U1}) in 
the context of a quantum dimer model on the honeycomb lattice. They found 
that the instanton fluctuations were irrelevant for $e'$ sufficiently 
small. Moreover, they found only one relevant higher order term - a cubic 
term $(\v E \cdot \v n_1) (\v E\cdot \v n_2) (\v E\cdot \v n_3)$. The 
same analysis can be applied in our case, but the cubic term is not allowed 
because of the symmetry $\v E \rightarrow -\v E$. 

Thus, the $z=2$ critical point (\ref{z2U1}) is potentially stable (just 
as in the previous section, we cannot rule out the possibility of a first 
order phase transition). This is an example of a 
deconfined quantum critical point. While the two adjoining phases differ 
by simple $Z_6$ symmetry breaking, the phase transition is not captured 
by a Landau-Ginzburg-Wilson action. Instead, the critical theory is 
described by gauge fluctuations which become deconfined only at the 
critical point. 

This deconfinement is physically reasonable - the critical 
point connects a liquid state of small strings to a phase with an ordered 
state made up of infinitely long strings. Thus it is natural that the 
transition point is described by a liquid state of long strings.
From this picture, one might expect the same phenomenon to occur in 
$(3+1)D$ spin models. In that case, one would expect an entire deconfined 
\emph{phase} between a small string phase and a striped phase. 

\section{Conclusion}

In this paper, we have described a mean field technique for quantum 
string (or dimer) models. The technique can be used to estimate 
phase diagrams, to analyze the low energy dynamics in each of the
phases, and to understand the critical points separating them.

The new mean field theory developed here is more powerful than traditional
mean field approaches in that it is applicable to both string condensed
phases and the usual symmetry breaking phases. Thus it can be used
to study phases and phase transitions beyond Landau's symmetry breaking
paradigm. One particularly interesting application is to frustrated 
spin systems with emergent photon-like excitations. 

We have demonstrated the approach with a simple example: the XXZ model 
(\ref{spkag}) in the limit $J_1 \gg J_{xy} \gg |J_1-J_2|$. In that 
limit, the low energy physics of the XXZ model is described by a
quantum string model which can be studied using the mean field theory. 
We find that the model is in a paramagnetic phase for large positive 
$J/g$, in a plaquette ordered phase for large negative $J/g$, and that 
the phase transition is in the 3D XY universality class. We have also 
applied the mean field approach to the XXZ model (\ref{spkag}) with an 
additional next-nearest neighbor coupling $J_{3}\sum_{\<\<\v i \v j\>\>} 
S^z_{\v i} S^z_{\v j}$. The mean field theory predicts that the model is 
in a paramagnetic phase for large positive $J/g$, in a stripe ordered 
phase for large negative $J/g$, and that the phase transition is a $z=2$ 
deconfined quantum critical point \eq{z2U1}.  

Given these results, it would interesting to study the 
XXZ model (\ref{spkag}) numerically. A quantum Monte Carlo study could 
potentially access the $z=2$ deconfined quantum critical point.
In addition, it could resolve the discrepancy between the resonating
plaquette phase predicted by \Ref{XM0555} and the frozen plaquette
phase that appears to be favored by the mean field approach (Fig. 
\ref{frozresplaq}).

A natural direction for future research would be to
apply the mean field approach to a $(3+1)$ dimensional spin model. 
Indeed, the mean field method may be most useful in this context. While
string condensed phases with emergent photons are always unstable in 
$(2+1)$ dimensions, there is no such problem in $(3+1)$ dimensions. Thus, 
the mean field method can be used to find \emph{entire phases} with 
string condensation and emergent photons, in addition to deconfined 
quantum critical points. 

\acknowledgments
We would like to thank Matthew Fisher, Leon Balents, Doron Bergman, and T. 
Senthil for useful discussions. This research was supported by NSF grant No.  
DMR-0433632 and ARO grant No. W911NF-05-1-0474.
\appendix

\section{Minimizing the energy}
\label{minenergy}
In general, the mean field phase diagram should be computed by minimizing 
the ground state energy $\<H\>$ over all choices for 
$\Psi_{\{z\}}$, and then identifying the quantum phase associated with the 
minimum energy $\Psi_{\{z\}}$. Ideally, the minimization of $\<H\>$ 
should be done in an unbiased fashion and include all possible $\{z\}$. 
However, to simplify our numerics, we have executed a more restricted 
minimization. In this section, we describe this minimization 
procedure.

In the restricted minimization procedure, we only minimize $\<H\>$ over 
the string liquid states, $z_{\v i \v j} = 
\alpha$. We then check to make sure that the minimal $z_{\v i \v j} = 
\alpha$ state is stable to infinitesimal perturbations $\alpha \rightarrow 
\alpha + \delta z_{\v i \v j}$.

To do this, we parameterize the perturbations by
$\alpha + \delta z_{\v i \v j} = \alpha e^{E_{\v i\v j} + i A_{\v i \v 
j}}$
where $E_{\v i \v j}$ and $A_{\v i \v j}$ are real - as in section 
\ref{lowen}. Expanding the energy to quadratic order in $E,A$, one finds
\begin{equation}
\<H\> = \sum_{\v i \v j, \v k \v l} (c_{\v i \v j, \v k \v l}
E_{\v i \v j} E_{\v k \v l} +
d_{\v i \v j, \v k \v l} A_{\v i \v j} A_{\v k \v l})
\label{posdef}
\end{equation}
where the constants $c_{\v i \v j, \v k \v l}, d_{\v i \v j, \v k \v l}$
are given by the equal time correlation functions (\ref{coeff}).

To check for stability one needs to check whether the matrices
$c_{\v i \v j, \v k \v l}, d_{\v i \v j, \v k \v l}$ are positive 
definite. This can be accomplished most easily by going to Fourier space, 
where $c_{\v i \v j, \v k \v l}, d_{\v i \v j, \v k \v l}$ are diagonal.
If all of the resulting eigenvalues are positive, then
the state $\Psi_{\alpha}$ is stable. Otherwise it is unstable.

If it is stable, we assume that it is the true lowest energy state. If it 
is unstable, say in the direction
$\alpha \rightarrow \alpha e^{E_{\v i\v j} + i A_{\v i \v j}}$, we 
conclude that the system enters a new (symmetry breaking) phase with an 
ordering given by $\alpha e^{E_{\v i\v j} + i A_{\v i \v j}}$. In effect, 
by restricting attention to local instabilities, we assume that the phase 
transitions out of the liquid phases are second order or weakly first 
order.

This restricted minimization procedure is less powerful and less reliable
then a general minimization of $\<H\>$ over all $\Psi_{\{z\}}$. Its only 
advantage is that it is technically simpler to implement.

\section{Calculation of mean field phase diagram}
\label{mfphdiag}

\begin{figure}
\centerline{
\includegraphics[width=3.0in]{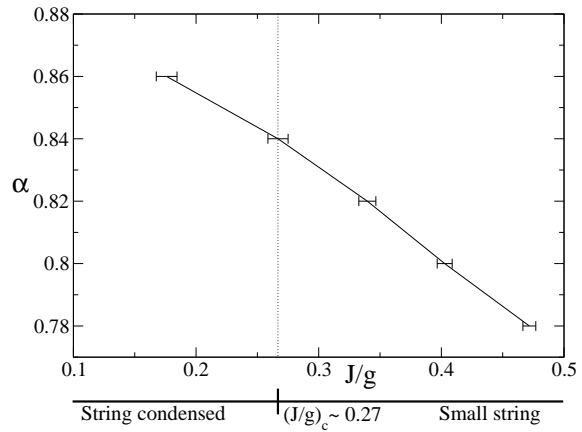}
}
\caption{
A plot of the minimum energy $\alpha$ as a function of $J/g$, obtained 
from a variational Monte Carlo simulation on a $12 \times 12$ lattice. The 
(mean field) critical point separating the string condensed and small
string phases is given by the value of $(J/g)$ where
$\alpha = \alpha_c \approx 0.84$.
}
\label{K12}
\end{figure}

In the following, we describe in detail how the mean field phase 
diagram for (\ref{Hstr}) was obtained. We begin with the transition at 
$J/g = 0.27$. The transition at $J/g = 0.27$ was obtained by calculating 
the optimal (minimal energy) $\alpha$ for different values of $J/g$. 
Plotting $\alpha$ as a function of $J/g$, we find that when $J/g$ is 
reduced below $0.27$, the minimum energy $\alpha$ becomes larger than 
$\alpha_c = 2^{-1/4} = 0.8409...$ (see Fig. \ref{K12}). Thus, at this 
point, the variational ground state enters the string condensed phase.

\begin{figure}
\centerline{
\includegraphics[width=3.0in]{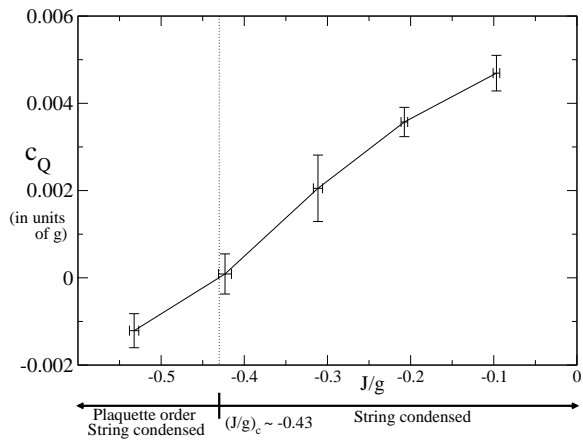}
}
\caption{
The eigenvalue $c_{\v Q}$ corresponding to the mode
(\ref{unstabmode}), as a function of $(J/g)$. The liquid becomes unstable
when $c_{\v Q}$ becomes negative. From the above plot, we find an
instability at $(J/g)_c \sim -0.43$.
}
\label{stab123}
\end{figure}

The string condensed (liquid) phase persists for $-0.43 < J/g < 0.27$. 
When $J/g < -0.43$, we find that the liquid becomes unstable
(see Fig. \ref{stab123}). In fact, two modes become unstable 
simultaneously. The two unstable modes are of the form
\begin{eqnarray}
E_{\v r \v s} &=& E_{\pm \v Q, \v s \v r} = \frac{1}{\sqrt{3Na^2}}(\v n 
\cdot (\v s
- \v r) e^{\pm i \v Q \cdot \v r}
\label{unstabmode}\\
A_{\v r \v s} &=& 0  \nonumber
\end{eqnarray}
where $\pm Q$ are the two wave vectors shown in Fig. \ref{brillouin}, and
$\v n \perp \v Q$.

Because two modes become unstable simultaneously, there is an ambiguity in
the way the system orders. The system could potentially order in any (real) 
linear combination of the two modes:
$E_{\v r \v s} = \zeta E_{+\v Q,\v r \v s} + \zeta^* E_{-\v Q,\v r \v s}$.

A purely quadratic analysis cannot distinguish between this continuum of
possible orderings. However, we expect that higher order terms will pick 
out a particular ordering. Indeed, let us consider the energy as a function
of the complex parameter $\zeta$, $\<H\> = H(\zeta)$. The symmetry of the
lattice requires that $H(\zeta) = H(-\zeta) = H(\zeta^*)$ and
$H(\zeta) = H(\omega \zeta)$ where $\omega$ is a third root of unity. The 
most general form for $H$ satisfying these constraints is
\begin{equation}
H(\zeta) = A |\zeta|^2+B |\zeta|^4+C (\zeta^6+(\zeta^*)^6)+...
\label{genzeta}
\end{equation}
to sixth order in $\zeta$. The first two terms are not sensitive to the 
phase of $\zeta$ and therefore tell us nothing about which linear 
combination is favored. However, the third term does pick out a phase. The 
phase depends on whether $C$ is positive or negative. If $C$ is positive, 
imaginary $\zeta$ is favored. This corresponds to a ``frozen 
plaquette'' phase - where a third of the plaquettes are typically occupied 
by strings, and the interstitial bonds are likely to be empty (Fig. 
\ref{frozresplaq}a). On the other hand, if $C$ is negative, real $\zeta$ is 
favored. This corresponds to a ``resonating plaquette'' phase where a 
third of the plaquettes resonate between two different configurations, 
while the interstitial bonds are typically occupied (Fig. 
\ref{frozresplaq}b).

While it is difficult to determine which of these two possibilities occur 
using the restricted minimization procedure, one can make a determination
by implementing a general minimization of $\<H\>$ over all 
$\Psi_{\{z\}}$. We have implemented this on a $3 \times 3$ lattice and the
result is that the frozen plaquette phase is favored. Based on this 
result, we believe that $C$ is positive and the 
instability at $J/g = -0.43$ results in frozen plaquette order. However,
a more complete numerical study is necessary to even make a definitive 
\emph{mean-field} prediction.

When $J/g$ is decreased below $-0.43$ the system acquires (frozen) 
plaquette order. However, the ordering is weak near the transition point, 
and a finite amount of ordering is necessary to destroy string 
condensation. This means that the string condensation persists for a finite 
interval below $-0.43$ - the system enters into a phase with 
\emph{simultaneous} plaquette order and string condensation.

We have not executed systematic numerics beyond $-0.43$. However, small
lattice results suggest that plaquette order strengthens as $J/g$ 
decreases. This suggests that when $J/g$ becomes sufficiently large and 
negative, the plaquette order becomes sufficiently strong that string 
condensation can no longer coexist and is destroyed. The system then enters 
a phase with the same plaquette order but no string condensation.

The phase diagram for the model (\ref{spkag}) with next nearest neighbor
coupling $J_{3}\sum_{\<\<\v i \v j\>\>} S^z_{\v i} S^z_{\v j}$ was 
computed using the same technique. We will not repeat the details here
because of their similarity to those described above.


\newcommand{\noopsort}[1]{} \newcommand{\printfirst}[2]{#1}
  \newcommand{\singleletter}[1]{#1} \newcommand{\switchargs}[2]{#2#1}

\end{document}